\documentclass{article}

\usepackage{PRIMEarxiv}

\usepackage[utf8]{inputenc} 
\usepackage[T1]{fontenc}    
\usepackage{hyperref}       
\usepackage{url}            
\usepackage{booktabs}       
\usepackage{amsfonts}       
\usepackage{nicefrac}       
\usepackage{microtype}      
\usepackage{lipsum}
\usepackage{fancyhdr}       
\usepackage{graphicx}       
\graphicspath{{media/}}     

\usepackage{multirow}

\title{Synthetic Misinformers: Generating and Combating Multimodal Misinformation
}

\usepackage{authblk}

\author[1, 2]{\small Stefanos-Iordanis Papadopoulos\thanks{Corresponding author} \ }
\author[1]{\small Christos Koutlis}
\author[1]{\small Symeon Papadopoulos}
\author[2]{\small Panagiotis C. Petrantonakis}

\affil[1]{\footnotesize Information Technology Insitute, Centre for Research \& Technology, Hellas.}
\affil[2]{\footnotesize School of Electrical \& Computer Engineering, Aristotle University of Thessaloniki.}
\affil[ ]{\textit {\{stefpapad,ckoutlis,papadop\}@iti.gr, \textit{ppetrant@ece.auth.gr}}}

\begin{document}
\maketitle

\begin{abstract}
With the expansion of social media and the increasing dissemination of multimedia content, the spread of misinformation has become a major concern.
This necessitates effective strategies for multimodal misinformation detection (MMD) that detect whether the combination of an image and its accompanying text could mislead or misinform.
Due to the data-intensive nature of deep neural networks and the labor-intensive process of manual annotation, researchers have been exploring various methods for automatically generating synthetic multimodal misinformation - which we refer to as Synthetic Misinformers - in order to train MMD models.
However, limited evaluation on real-world misinformation and a lack of comparisons with other Synthetic Misinformers makes difficult to assess progress in the field.
To address this, we perform a comparative study on existing and new Synthetic Misinformers that involves (1) out-of-context (OOC) image-caption pairs, (2) cross-modal named entity inconsistency (NEI) as well as (3) hybrid approaches and we evaluate them against real-world misinformation; using the COSMOS benchmark. 
The comparative study showed that our proposed \textit{CLIP-based Named Entity Swapping} can lead to MMD models that surpass other OOC and NEI Misinformers in terms of multimodal accuracy and that hybrid approaches can lead to even higher detection accuracy. 
Nevertheless, after alleviating \textit{information leakage} from the COSMOS evaluation protocol, low Sensitivity scores indicate that the task is significantly more challenging than previous studies suggested. Finally, our findings showed that NEI-based Synthetic Misinformers tend to suffer from a unimodal bias, where text-only MMDs can outperform multimodal ones.

\end{abstract}


\section{Introduction}

The proliferation of misinformation is a growing challenge in today's society, especially with the widespread use of social media and the Internet. 
Consequently, the automatic detection of misinformation has become an important challenge, with researchers exploring various methods for identifying false claims through natural language processing \cite{mridha2021comprehensive} and detecting manipulated images, such as DeepFakes, through computer vision techniques \cite{rana2022deepfake}. 
The aforementioned challenges primarily focus on individual modalities, necessitating the use of unimodal detection models. However, multimedia content has been shown to be more attention-grabbing and widely disseminated than plain text content \cite{li2020picture}. 
Furthermore, the presence of an image can make a false statement more convincing to individuals \cite{newman2012nonprobative}, emphasizing the importance of multimodal misinformation detection (MMD).

\begin{figure}[!ht]
    \centering
    \includegraphics[width=0.8\textwidth]{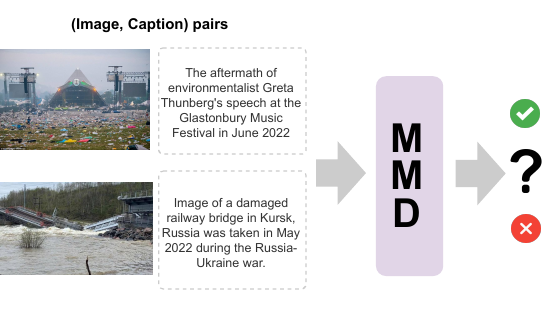}
    \caption{Multimodal misinformation detection (MMD) models attempt to identify whether an (Image, Caption) pair is 
    truthful 
    or misleading.
    The images and captions are taken from reuters.com.
    }
    \label{fig:mmd}
\end{figure}

MMD models, as seen in Fig. \ref{fig:mmd}, are trained to identify whether an image and its accompanying caption in combination are accurate (truthful) or misleading (leading to misinformation).
The image at the top of Fig. \ref{fig:mmd} shows a music festival with littered grounds, while the accompanying caption claims that the event occurred in June 2022, after a speech by environmentalist Greta Thunberg.  
In reality, even though the image depicts a Glastonbury festival, it was not taken in June 2022 after Greta Thunberg’s speech, but rather in 2015; years before Greta Thunberg became a public figure\footnote{\url{https://www.reuters.com/article/factcheck-glastonbury-greta-idUSL1N2YE1JD}}.
This case involves the manipulation of named entities (person and date) in order to frame a target public figure and her associated audience under a negative spotlight.
The bottom image illustrates a railway bridge that collapsed into a body of water with the caption claiming that the event took place in Kursk, Russia during the 2022 Russia-Ukraine war. 
A bridge was indeed damaged that day in Kursk, however, this image was actually taken in 2020 in Murmansk, Russia\footnote{\url{https://www.reuters.com/article/factcheck-destroyed-bridge-idUSL2N2WU1CM}}. 
This case illustrates the use of an out-of-context image, either by mistake or with the intention to exaggerate or downplay the severity of the event. 
From the above examples, it is evident that multimodal misinformation takes diverse forms, can be disseminated for various motives, and entails subtle cues that are challenging to discern.

\begin{figure*}[!ht]
    \centering
    \includegraphics[width=1\textwidth]{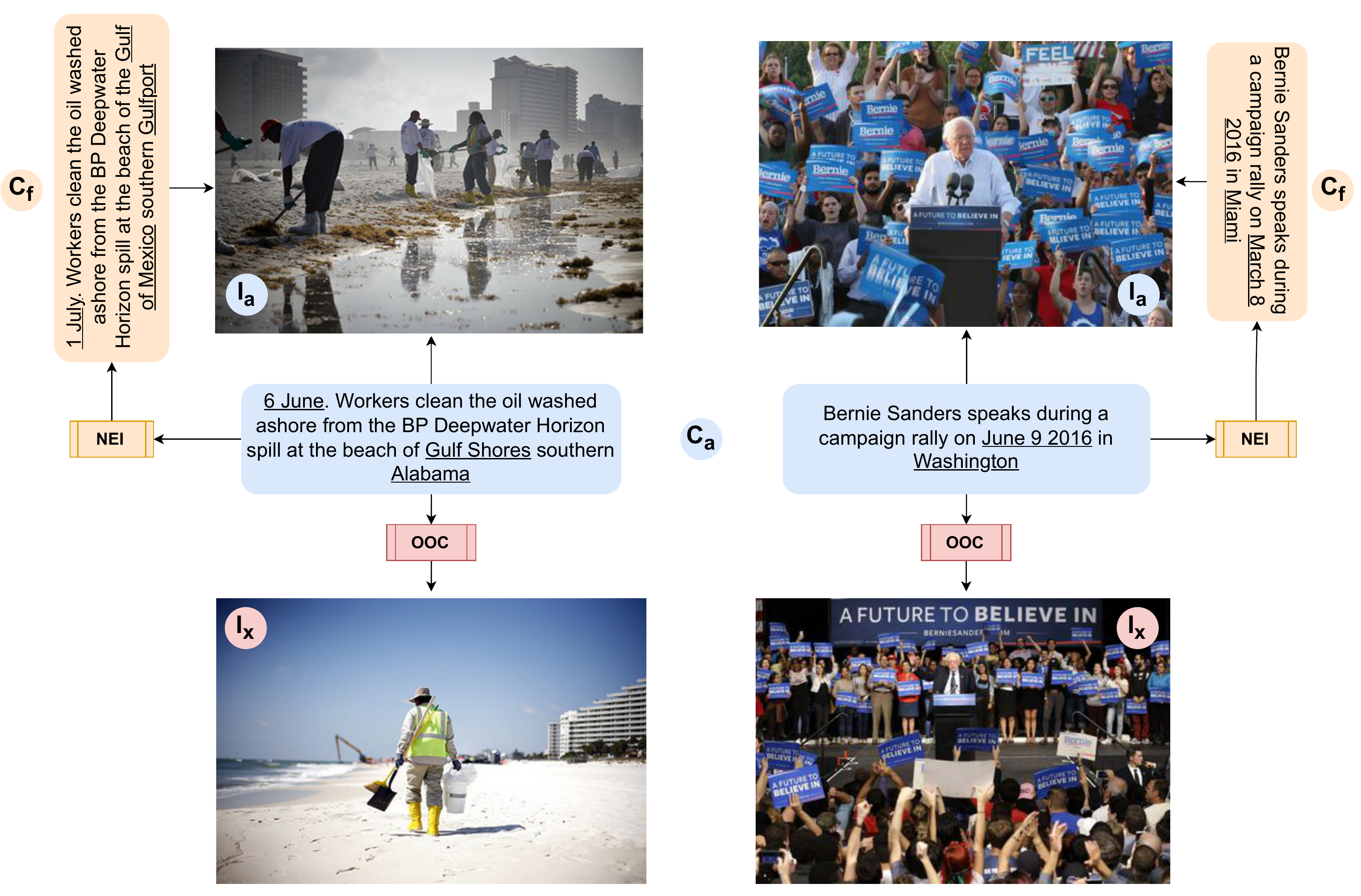}
    \caption{Training data generated by two types of Synthetic Misinformers: one creates out-of-context (OOC) misinformation and the other produces cross-modal named entity inconsistencies (NEI).
    Given a truthful ($I_a, C_a$) image-caption pair, OOC samples an image $I_x$ creating ($I_x, C_a$) 
    while NEI manipulates the named entities in $C_a$ to create the falsified $C_f$.
    These examples were generated by CLIP-based sampling (\textit{CSt-alt}) and CLIP-based named entity swapping (\textit{CLIP-NESt-alt}) for OOC and NEI respectively (See Section \ref{sec:misinformers} for more details on these methods). 
    The images and captions are taken from the VisualNews dataset \cite{liu2020visual}. 
    }
    \label{fig:samples}
\end{figure*}

Researchers have been exploring training deep neural networks for MMD \cite{alam2021survey}. 
Considering the data-intensive nature of training deep neural networks for MMD as well as the time-consuming and labor-intensive nature of manual annotation, researchers have been investigating methods for automatically generating synthetic multimodal misinformation, which we refer to as Synthetic Misinformers. 
These methods include the generation of out-of-context (OOC) image-text pairs or the creation of cross-modal entity inconsistencies (NEI).
OOC involves 
pairing an image with an incongruous caption \cite{aneja2021cosmos} while NEI involves manipulating the named entities in otherwise truthful captions \cite{muller2020multimodal}.
Previous works have relied on random sampling 
\cite{jaiswal2017multimedia, aneja2021cosmos} or feature-informed sampling methods \cite{luo2021newsclippings, biamby2021twitter} for generating OOC and in-cluster random sampling \cite{sabir2018deep} or rule-based random sampling \cite{muller2020multimodal} for generating NEI. Examples of generated OOC and NEI misinformation can be seen in Fig. \ref{fig:samples}. 

\begin{figure*}[!ht]
    \centering
    \includegraphics[width=1\textwidth]{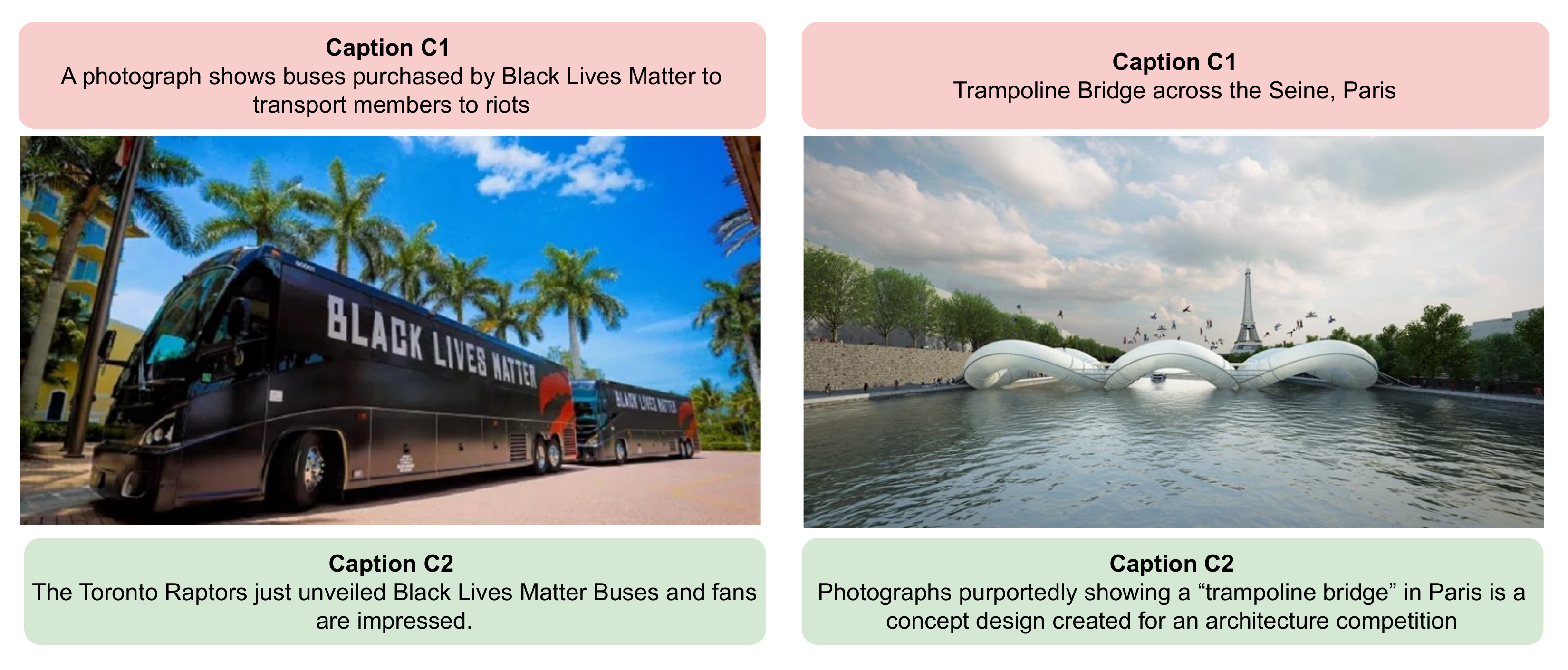}
    \caption{Examples of falsified $(I, C1, C2)$ triplets from the COSMOS benchmark \cite{aneja2021cosmos}.
    Caption $C2$ either provides the correct description for the image (left) or an explanation of why $C1$ is false (right).
    We consider $C2$ to be \textit{information leakage} and excluded them from the evaluation protocol.}
    \label{fig:cosmos_samples}
\end{figure*}

Nevertheless, despite the efforts of previous studies, those have been limited by evaluating their methods on test sets generated by their own Synthetic Misinformers instead of real-world multimodal misinformation,
the only exception being the work presenting the COSMOS dataset \cite{aneja2021cosmos}. 
However, the authors of the latter work made use of a problematic evaluation protocol that suffers from \textit{information leakage}, as seen in Fig.\ref{fig:cosmos_samples} (See section \ref{sec:eval_protocol} for further details).
Additionally, prior studies did not compare their methods with other Synthetic Misinformers.
Lack of comparison and evaluation on real-world data hinders the ability of the research community to assess the progress made and determine the current state-of-the-art on MMD.
To this end, we replicate multiple Synthetic Misinformers, 
fine-tune our Transformer-based MMD model (termed DT-Transformer) on the generated data and finally compare them on the COSMOS benchmark \cite{aneja2021cosmos} that encompasses real-world multimodal misinformation.
In our comparative study, we examine whether OOC or NEI are more representative  of real-world misinformation. Our hypothesis is that -  despite prior research treating them as separate tasks - both OOC and NEI are crucial components of effective MMD. 
For this reason, we also investigate a range of hybrid Synthetic Misinformers.

The main contributions of our work can be summarised as follows:

\begin{itemize}

\item We perform the first comparative study on Synthetic Misinformers, which offers a comprehensive evaluation of the current state-of-the-art on MMD and can provide guidance for future research in the field.

\item We introduce ``CLIP-based Named Entity Swapping'', which demonstrates the highest multimodal accuracy among other OOC and NEI approaches. 
Additionally, we demonstrate that hybrid Synthetic Misinformers can further enhance detection accuracy.

\item Our findings highlight that although NEI-based may outperform OOC methods, they tend to suffer from a unimodal bias, where text-only models can outperform multimodal ones. 
Moreover, low Sensitivity 
scores
- or Hit Rate for `Falsified' pairs -
indicate that the problem is significantly more challenging than previous works suggested.
We offer recommendations on how future studies may address these challenges.

\end{itemize}

\section{Related Work}

Studies on multimodal misinformation detection (MMD) have focused on out-of-context image-language pairs (OOCs) or cross-modal named entity inconsistencies (NEI). 
One common form of multimodal misinformation involves de-contextualization; a legitimate image being paired with an out-of-context caption creating a deceptive impression. Consequently, researchers have used random-sampling (\cite{jaiswal2017multimedia, aneja2021cosmos}) and feature-informed sampling methods (\cite{luo2021newsclippings, biamby2021twitter}) for generating OOCs. 
The MAIM dataset was created by randomly sampling image-text pairs collected among image-caption pairs collected from Flickr \cite{jaiswal2017multimedia}.
The authors developed a joint embedding with the use of deep representation learning and then calculated the image-caption consistency. 
Similarly, Aneja et al.\cite{aneja2021cosmos} created the COSMOS training dataset by collecting truthful image-captions pairs from credible news websites and then matching captions with random images to create OOCs. 
The authors utilized self-supervised deep learning and evaluated their method on the COSMOS benchmark, consisting of real-world multimodal misinformation.
However, random sampling can not ensure that the image-caption pair will bear any relation and tends to generate easy negative samples that do not resemble realistic multimodal misinformation ; capable of deceiving humans. 
To this end, Luo et al.\cite{luo2021newsclippings} created the NewsCLIPings datasets by utilizing the large cross-modal CLIP model \cite{radford2021learning} along with scene-learning and person matching models in order to generate hard negative samples. 
Similarly, the Twitter-COMMs dataset was created by combining and applying CLIP-based sampling (to generate hard negatives) and in-topic random sampling (to resolve class imbalance) on data collected from Twitter\footnote{\url{https://twitter.com}}, related to three topics: climate, COVID, and military vehicles \cite{biamby2021twitter}.

On the other hand, NEI involves legitimate images being accompanied by a manipulated caption whose named entities (person, location, event) do not match with the content or the context of the image. 
The ``Multimodal Entity Image Re-purposing'' (MEIR) dataset was created by clustering image-caption pairs based on `relatedness' - location proximity, text and image similarity - by using GPS coordinates, word2vec and VGG19 pre-trained on ImageNet respectively. 
Then they randomly swap named entities of the same type between the current caption and another caption taken from the same cluster \cite{sabir2018deep}. 
Similarly, the TamperedNews dataset was created by randomly replacing named entities with ones of the same type given that the replaced person is of the same gender and/or country, locations are within high geographical proximity and events belong to the same category (e.g sport competitions or 
natural disasters) \cite{muller2020multimodal}. 

The above works either provide internal ablation \cite{sabir2018deep, muller2020multimodal, luo2021newsclippings, biamby2021twitter}
or comparison with simple baselines \cite{sabir2018deep}
and do not compare their methods with other Synthetic Misinformers.
Moreover, studies that used some of the above datasets have mostly focused on incremental methodological improvements \cite{akgul2021cosmos}, such as the integration of sentiment \cite{alkaddour2022sentiment} or evidence \cite{abdelnabi2022open} in MMD models. 
With the exception of \cite{aneja2021cosmos}, prior works have not evaluated their methods on real-world misinformation but on data generated by their own Synthetic Misinformer.
Therefore, there is no clear way for the research community to assess progress in the field including the current state-of-the-art on MMD, which s the best way to generate training data for MMD and whether OOC or NEI (or both) are better representative of real-world misinformation.
To address this gap, we perform a comparative study on various Synthetic Misinformers - both OOC, NEI and hybrid methods - and evaluate them on real-world multimodal misinformation. 

Finally, Luo et al.\cite{luo2021newsclippings} argued that methods utilizing named entity manipulations may introduce linguistic biases. 
To investigate this, the authors trained a text-only BERT \cite{devlin2018bert} model and achieved similar results to the multimodal models used in \cite{muller2020multimodal}. 
However, the latter extracted visual features from an ImageNet pre-trained ResNet and textual features from off-the-self fastText \cite{bojanowski2017enriching}. 
At the time of writing (2021 \cite{luo2021newsclippings}), BERT-like models were considered among the state-of-the-art for text-based tasks while fastText was an older architecture, rendering the training protocols significantly different and thus not directly comparable. Therefore, it is not possible to conclude definitively about the existence of unimodal bias based solely on these results. To address this gap, we re-examine whether NEI methods suffer from unimodal bias, within a controlled training framework and an evaluation on real-world misinformation.  

\section{Methodological Framework}

\subsection{Problem Formulation}

\begin{figure*}[!ht]
    \centering
    \includegraphics[width=1\textwidth]{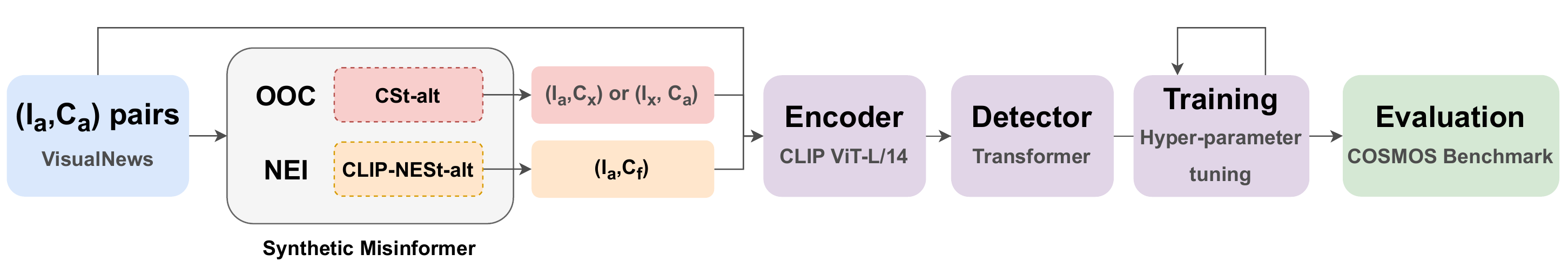}
    \caption{High-level overview of the proposed workflow. 
    Truthful ($I_a, C_a$) image-caption pairs are manipulated by a Synthetic Misinformer which generates Falsified pairs. Truthful and Falsified pairs are encoded by CLIP ViT-L/14 and used to train the Detector who is optimized for binary classification when using one OOC or NEI dataset, or for multi-class classification when combining one OOC and one NEI method. 
    Here, we showcase the hybrid Synthetic Misinformer \textit{CLIP-NESt-alt + CSt-alt}.}
    \label{fig:pipeline}
\end{figure*}

In this study we compare numerous Synthetic Misinformers, methods for generating synthetic multimodal misinformation, 
and evaluate them on real-world multimodal misinformation. 
The problem is defined as a binary classification task, where an (I,C) image-caption pair is either truthful or falsified. 
Truthful captions are collected from credible news sources while falsified ones are produced by a Synthetic Misinformer. 
Each (I,C) is encoded by a visual encoder
E\textsubscript{V}($\cdot$) and a textual encoder E\textsubscript{T}($\cdot$) that produce the corresponding vector representations $\mathbf{v}_{I}\in\mathbb{R}^{e\times 1}$ and $\mathbf{t}_{C}\in\mathbb{R}^{e\times 1}$ for the image $I$ and caption $C$ respectively, where $e$ is the encoder's embedding dimension. 
The extracted features are concatenated and passed through the multimodal detection deep learning neural network D($\cdot$) - referred to as the Detector - whose parameters have to be optimized and its hyper-parameters tuned. Finally, the predictions of the trained Detector will be evaluated against a test set consisting of real-world multimodal misinformation.
In order to accurately and fairly compare various Misinformers, they need to share a common training and evaluation framework. 
Therefore, the (1) Encoder, (2) Detector, (3) optimization and hyper-parameter tuning and the (4) evaluation process should remain constant during the comparative study while only the Misinformer method changes.
This framework ensures that any change in performance will be the result of the Misinformer and not other factors. A high-level illustration of the proposed workflow can be seen in Fig. \ref{fig:pipeline}.
In this section, we address each aspect individually.

\subsubsection{\textbf{Encoder}}

After generating a training dataset with a Synthetic Misinformer, we use an Encoder for extracting visual and textual features from all images and captions that will be used to train the Detector. 
The first works on MMD mostly relied on convolutional neural networks pre-trained on ImageNet to extract features from images (namely VGG-19 \cite{jaiswal2017multimedia, sabir2018deep} and ResNet50 \cite{muller2020multimodal, aneja2021cosmos}) and word embeddings to extract features from captions (namely word2vec \cite{jaiswal2017multimedia, sabir2018deep} and fastText \cite{muller2020multimodal}).
More recent approaches, have utilized large-scale multimodal and cross-modal models, namely CLIP \cite{luo2021newsclippings, biamby2021twitter, huang2022text}, VisualBERT \cite{luo2021newsclippings} and VinVL \cite{huang2022text} to extract both their visual and textual features. 
In the aforementioned works, CLIP \cite{radford2021learning} tended to outperform other cross-modal methods (VinVL and VisualBERT) for MMD \cite{luo2021newsclippings, biamby2021twitter, huang2022text}.

Contrastive language-image pre-training, or CLIP in short, is a cross-modal model trained to match the most relevant text to an image. Developed by Radford et al.\cite{radford2021learning} and trained on a large-scale dataset -  approximately $4\times10^8 $ image-text pairs - CLIP has proven to have powerful zero-shot capabilities; meaning that it performs well on tasks and domains that it was not explicitly trained for.
In our study, we first perform an experiment using CLIP ViT-B/32 on the NewsCLIPings datasets, in order to compare our training pipeline with \cite{luo2021newsclippings} 
but we also utilize the updated and improved\footnote{\url{https://huggingface.co/sentence-transformers/clip-ViT-L-14}} CLIP ViT-L/14 version in the comparative study.
CLIP ViT-B/32 produces an embeddings vector of size $e = 512$ while ViT-L/14 produces $e=768 $.  
We use CLIP off-the-shelf\footnote{\url{https://github.com/openai/CLIP}} and do not fine-tune it further 
due to computational resource constraints. 
Luo et al.\cite{luo2021newsclippings} experimented with fine-tuning the whole or the top layers of CLIP-ResNet-50, but their results were mixed; fine-tuning could not consistently outperform the ``frozen'', off-the-shelf, CLIP in all cases. There have been proposed methods for robustly fine-tuning large-scale cross-modal neural networks \cite{wortsman2022robust} but they are outside the scope of this study; since we do not attempt to reach the highest possible performance but primarily focus on providing a fair comparative study of various Synthetic Misinformers. 

\subsubsection{\textbf{Detector}}

Previous works either calculate a cross-modal similarity score \cite{jaiswal2017multimedia, muller2020multimodal, aneja2021cosmos, akgul2021cosmos} or define binary classifier on top \cite{luo2021newsclippings, alkaddour2022sentiment, sabir2018deep, biamby2021twitter}. 
Few works have also added fully-connected layers to analyse the extracted features before the final classification layer \cite{ alkaddour2022sentiment, sabir2018deep}. 
Instead, we consider the Transformer architecture \cite{vaswani2017attention} is an even more appropriate choice for the Detector.
Our Transformer-based Detector (DT-Transformer in short), first concatenates the encoded captions and images and pass them through a Transformer architecture consisting of $L$ layers that have $h$ attention heads of embedding dimension $d$.
Its output is then passed through a normalization layer, a dropout layer, a fully connected layer which is activated by the GELU function, a second dropout layer and a final binary classification layer. 

\subsubsection{\textbf{Optimization}}
Given that we define MMD as a binary classification task, the Detector is optimized based on the binary cross entropy loss function. 
Due to differences in distribution, scale, complexity and other factors, we assume that the Detector may require different hyper-parameters to perform optimally with different training datasets. 
To this end, we tune the Detector's hyper-parameters based on the following grid search: $L \in \{1, 4\}$ transformer layers of $d \in \{128, 1024\}$ dimensions, $h \in \{2, 8\}$ attention heads, and a learning rate of $lr \in \{1e-4, 5e-5\}$. The dropout rate is constant at 0.1 and the batch size at 512. 
The selected hyper-parameter grid amounts to 16 experiments for each Synthetic Misinformer dataset. This is clearly not exhaustive, but  adding any more options would exponentially increase the required time and computational resources. 
Instead, our aim is to give the chance to each method to reach an adequate and representative performance, even if it is not the globally optimal that would be possible through exhaustive optimization. 
The Detector is optimized by the ADAM optimizer for a maximum of 30 epoch with early-stopping at 10 epochs. 
At the end, we retrieve the checkpoint with the highest validation accuracy and use it for the final evaluation on the test set. 

\subsubsection{\textbf{Evaluation protocol}}
\label{sec:eval_protocol}

With the exception of Aneja et al.\cite{aneja2021cosmos}, previous works on Synthetic Misinformers have not evaluated their methods on real-world misinformation. 
Instead, 
they first propose a method for generating multimodal misinformation which they apply on a body of truthful image-captions pairs. After generating the dataset, the authors split it into training, validation and test sets and report the best performance on the test set. 
This only tells us how a model trained on a synthetic dataset will perform on a dataset generated by the same process, not how accurately it could potentially detect misinformation ``in the wild''. 
Moreover, prior works have not provided direct comparison with other Synthetic Misinformers and therefore, we can not assess progress or the current state-of-the-art in the field. 
To the best of our knowledge, COSMOS \cite{aneja2021cosmos} is the only publicly available, manually annotated benchmark for MMD\footnote{The DARPA SemaFor is another benchmark consisting of real-world misinformation and was used in \cite{biamby2021twitter, huang2022text} - consisting of 200 falsified instances - but we were unable to obtain access.}.  
The COSMOS benchmark 
- also used in the MMSys’21 ``Grand Challenge on Detecting Cheapfakes'' \cite{aneja2021mmsys} - consists of 850 image-caption pairs from the fact-checking website SNOPES\footnote{\url{https://www.snopes.com/}} and 850 truthful image-caption pairs from credible news sources. 
Therefore, we will use the COSMOS evaluation set for our study. 

Nevertheless, it is important to highlight certain problematic aspects of the evaluation protocol used in \cite{aneja2021cosmos}.
During evaluation, the authors provide a triplet of an image and two captions $(I, C1, C2)$
and make a threshold-based decision by examining $C1$-$C2$ similarity and their overlap with the objects in the image. 
First, this protocol does not reflect how we encounter real-world misinformation, with an image usually being accompanied by a single caption or a small paragraph (e.g on Twitter).
More importantly, $C2$ in falsified instances is either an explanation of why $C1$ is false or the truthful caption for the image. 
In two examples taken from the COSMOS benchmark and shown in Fig. \ref{fig:cosmos_samples}, $C2$ reads 
``Toronto Raptors just unveiled Black Lives Matter Buses and fans are impressed'' and 
``Photograph showing a `trampoline bridge' in Paris is a concept design for an architecture competition''.
This is a clear case of \textit{information leakage} and does not reflect how we encounter misinformation in the real world. 
Fact checkers are not usually presented with two separate bodies of texts and have to decide which is the correct one. 
Instead, they are usually presented with an image and a single body of text and they have to determine whether the text is truthful and whether the text accurately matches and describes the image.
It is important to note that these are not outlier cases. We have manually examined hundreds from the COSMOS benchmark and Caption 2 suffers from the same problem.
Therefore, we do not consider the 88\% detection accuracy reported by the authors to be representative \cite{aneja2021cosmos}. 
For that reason, in this study, we only use the (I,C1) tuples from the COSMOS benchmark. 

\subsection{Synthetic Misinformers
}
\label{sec:misinformers}

We define three types of Synthetic Misinformer, namely methods that generate: (1) out-of-context (OOC) image-captions pairs, (2) cross-modal entity inconsistency (NEI) where certain named entities in the caption are tampered and do not correspond with the content of the image, and (3) hybrid approaches that combine both OOC and NEI misinformation. 
Table  \ref{tab:datasets} displays the number of training samples produced by each Synthetic Misinformer (OOC or NEI) as well as the number of Truthful pairs.

\begin{table}
\centering
  \caption{Number of instances per class for training datasets generated by different Synthetic Misinformers. }
  
  \label{tab:datasets}
  \begin{tabular}{llll}
    \toprule

     \multicolumn{1}{l}{\textbf{Synthetic Misinformer}} &      
     \multicolumn{1}{l}{\textbf{Truthful}} &
     \multicolumn{1}{l}{\textbf{OOC}} & 
     \multicolumn{1}{l}{\textbf{NEI}} 
     \\
     \midrule

     VisualNews-Training & 1,007,785 & - & - \\
    \midrule
     NC/Bal & 35,536 & 35,536 & - \\

     NC/I-T & 226,564 & 226,564 & - \\ 
     
    NC/T-T & 258,036 & 258,036 & - \\         
     
     Random sampling & 1,007,744 & 1,007,744 & - \\
     CLIP-based sampling & 1,007,744 & 1,007,744 & - \\

     \midrule

     MEIR & 82,156 & - & 57,940 \\
     R-NESt & 1,007,744 & - & 924,586 \\
     CLIP-NESt-C & 1007744 & - & 835,537 \\
     CLIP-NESt-I & 1007744 & - & 859,618 \\
     CLIP-NESt-alt & 1007744 & - & 847,693 \\

     \midrule
     R-NESt + NC/I-T & 226,564 & 226,564 & 226,564\\
     R-NESt + CSt-alt & 1,007,744 & 1,007,744 & 924,586 \\
     CLIP-NESt-alt + CSt-alt & 1,007,744 & 982,645 & 847,693 \\

    \bottomrule
  \end{tabular}
\end{table}

\subsubsection{\textbf{Out-of-context misinformation}}

In order to create out-of-context (OOC) image-caption pairs, we first need  a dataset of truthful pairs ($I_a, C_a$) and then a method for sampling an OOC image $I_x$ or an OOC caption $C_x$. 
In this study we make use of the VisualNews dataset \cite{liu2020visual} that consists of 1,259,732 truthful $(I_a,C_a)$ pairs collected by four credible sources (The Washington Post, USA Today, The Guardian and the BBC) regarding 159 topics, namely: art and culture, world, law and crime, international relations, science and technology sports, environment, elections and others. 
We use the VisualNews training set to generate training data and the VisualNews validation set to generate the validation data in order to avoid overlapping samples and \textit{information leakage}. 
We experiment with the following OOC Synthetic Misinformer methods: 
\begin{itemize}
    \item Random sampling by caption (\textbf{RS-C}): for every actual ($I_a,C_a$) pair, sample a random caption $C_x$ from the whole corpus. This process was used to generate the COSMOS training set\cite{aneja2021cosmos} but we apply it on the VisualNews dataset instead. 

    \item In-topic random sampling by caption (\textbf{RSt-C}): for every ($I_a,C_a$), sample a random caption $C_x$ of the same topic as $C_a$ (e.g. international politics, elections, environment, etc). Using candidates from the same topic can increase the chance of relevance. 
    A similar process was used in \cite{biamby2021twitter} but only as a means to mitigate class imbalance. 
    We also define in-topic random sampling by image (\textbf{RSt-I}): sampling a random image $I_x$ for an actual ($I_a,C_a$) 
    and in-topic random sampling by alternating between image or caption (\textbf{RSt-alt}): choose whether to sample an $I_x$ or an $C_x$ at random (determined by a random selection function). 

    \item Similarly, we implement in-topic CLIP-based sampling by caption to caption similarity (\textbf{CSt-C}), by image-image similarity (\textbf{CSt-I}) or by alternating between image-image and caption-caption similarity (\textbf{CSt-alt}): calculate the most similar item ($I_x$ or $C_x$) based on features extracted from CLIP. As candidates we define $(I_a,C_a)$ pairs from the same topic and use the cosine similarity as the metric of similarity.  

    \item Finally, we experiment with three versions of the NewsCLIPings (NC) datasets, namely (1) NewsCLIPings Semantics / CLIP Text-Image \textbf{(NC/T-I}), (2) NewsCLIPings Semantics / CLIP Text-Text (\textbf{NC/T-T}) and the (3) NewsCLIPings Merged / Balanced (\textbf{NC/Bal}) as provided by the authors\footnote{\url{https://github.com/g-luo/news_clippings}} \cite{luo2021newsclippings}. 

\end{itemize}

\subsubsection{\textbf{Cross-modal named entity inconsistency}}
\noindent
In order to generate image-caption pairs that suffer from cross-modal named entity inconsistency we need truthful ($I_a,C_a$) pairs, a method for sampling and swapping the named entities in $C_a$ in order to create the falsified $C_f$. 
We use the VisualNews and experiment with the following methods: 
\begin{itemize}
    \item In-topic random named entity swapping (\textbf{R-NESt}): for every ($I_a,C_a$), identify all entities in $C_a$ and replace them with randomly sampled entities of the same type (person, location, organization, date, event etc) that belong to the same topic as $C_a$. 
    \item The MEIR dataset as was provided by the authors\footnote{\url{https://github.com/Ekraam/MEIR}}, but we extract features from CLIP ViT L/14 instead of using VGG19 as in the original paper \cite{sabir2018deep}. 
    \item We propose in-topic CLIP-based named entity swapping (CLIP-NESt) by image-image similarity (\textbf{CLIP-NESt-I}), caption-caption similarity (\textbf{CLIP-NESt-C}) or alternating between image-image and caption-caption similarity (\textbf{CLIP-NESt-alt}): for every ($I_a, C_a$) pair, identify the most similar ($I_x, C_x$) pair based on features extracted from CLIP and swap the entities of the same type between $C_a$ and $C_x$ in order to create $C_f$. 
    $C_x$ should have at least one named entity of the same type as $C_a$ but be a different named entity (avoid swapping one named entity with itself), otherwise we select the next most similar pair. 
    As candidates we define image-caption pairs from the same topic and use the cosine similarity as the metric of similarity. 
    We use the SpaCy Named Entity Recognizer (NER) and specifically the \texttt{en\_core\_web\_trf} module, which exhibits 0.90 F1-score for NER\footnote{\url{https://spacy.io/models/en\#en_core_web_trf-accuracy}}. 
    Our rationale for proposing CLIP-NESt, is that CLIP-based similarity will retrieve a semantically or thematically similar caption and as a result, their entities are more likely to be related in some aspect. Therefore, swapping entities between similar captions will create more plausible misinformation than randomly sampled ones.

\end{itemize}

\subsubsection{\textbf{Hybrid methods}}
We also experiment with methods that combine both OOC and NEI misinformation, which we refer to as hybrid methods. We follow the same training process but instead of binary classification, we train the Detector for multi-class classification with the use of the cross-entropy loss function. 
The Detector is trained to classify $(I, C)$ pairs into three classes: \texttt{Truthful}, \texttt{NEI} or \texttt{OOC}.
During evaluation on the COSMOS dataset, \texttt{NEI} and \texttt{OOC} predictions are set to \texttt{Falsified} pairs; since COSMOS is a binary dataset. 
In this study, we combine few of the best performing methods: (1) R-NESt + CSt-alt, (2) R-NESt + NC/I-T and (3) CLIP-NESt-alt + CSt-alt. 
Our motivation for exploring hybrid methods is that real-world multimodal misinformation may not be adequately represented by OOC or NEI alone, but may require a combination of both.
OOC and MEI methods produce balanced datasets, since they create one falsified pair for every truthful pair. 
On the other hand, some hybrid methods showcase imbalanced classes, in this case, we apply random down-sampling.

\section{Experimental Results}

Before starting with the comparative study, we wanted to examine whether our training pipeline - the choice of Encoder and Detector - was valid; if it outperforms or at the very least competes with previous works. 
To this end, we compare our training pipeline with \cite{luo2021newsclippings}. In Table \ref{tab:newsclippings} we observe that while using the same Encoder (CLIP ViT-B/32), our DT-Transformer can consistently outperform \cite{luo2021newsclippings} on all three NewsCLIPings datasets. 
Moreover, using the CLIP ViT-L/14 Encoder can significantly surpass ViT-B/32. We therefore proceeded our comparative study with the DT-Transformer and the CLIP ViT-L/14 Encoder. 

\begin{table*}
\centering
  \caption{Comparison between different Detectors and Encoders trained and evaluated on three NewsCLIPings datasets. We report the binary accuracy and the per-class recall for truthful and falsified image-caption pairs. 
  Our proposed Transformer-based Detector (DT-Transformer) consistently outperforms the original NewsCLIPings Detector and using features from CLIP ViT-L/14 further improves performance.
  We report the overall Accuracy and Hit Rate per class (Truthful and Falsified pairs) or Specificity and Sensitivity respectively.
  }
  
  \label{tab:newsclippings}
  \begin{tabular}{llllll}
    \toprule

     \multicolumn{1}{l}{\textbf{Synthetic Misinformer}} & 
     \multicolumn{1}{l}{\textbf{Detector}} &      
     \multicolumn{1}{l}{\textbf{Encoder}} & 
     \multicolumn{1}{l}{\textbf{Accuracy}} & 
     \multicolumn{1}{l}{\textbf{Truthful}} &
     \multicolumn{1}{l}{\textbf{Falsified}} 
     \\
     \midrule

     NC/Bal & \cite{luo2021newsclippings} & CLIP ViT-B/32 & 60.23 & 70.07 & 50.39 \\
     NC/Bal & DT-Transformer & CLIP ViT-B/32 & 65.67 & 73.73 & 57.60 \\
     NC/Bal & DT-Transformer & CLIP ViT-L/14 & 77.09 & 78.58 & 75.61 \\

    \midrule

     NC/T-T & \cite{luo2021newsclippings} & CLIP ViT-B/32 & 69.39 & 74.09 & 64.69 \\
     NC/T-T & DT-Transformer & CLIP ViT-B/32 & 71.63 & 75.56 & 67.69 \\
     NC/T-T & DT-Transformer & CLIP ViT-L/14 & 84.12 & 87.45 & 80.79 \\   

     \midrule

     NC/I-T & \cite{luo2021newsclippings} & CLIP ViT-B/32 & 66.98 & 75.43 & 58.53 \\
     
     NC/I-T & DT-Transformer & CLIP ViT-B/32 & 69.12 & 67.51 & 70.73 \\
     NC/I-T & DT-Transformer & CLIP ViT-L/14 & 80.98 & 84.62 & 77.33 \\           
     
    \bottomrule
  \end{tabular}
\end{table*}

Table \ref{tab:comparative} shows the comparative study between various Synthetic Misinformers. 
Commencing with the OOC Misinformers, we observe that in-topic candidates can improve random sampling (+3\% improvement). Since image-caption pairs are more likely to be related if they are taken from the same topic (e.g election) than completely random topics (e.g elections and sports). 
Secondly, we observe that alternating between sampling $(I_a, C_x)$ and $(I_x, C_a)$ pairs improves the performance of \textit{RSt}, and \textit{CSt}. 
Previous works on OOC Misinformers were only sampling $C_x$ captions. 
As a result, each image would appear twice in the dataset while the captions could appear once, twice or multiple times; depending on the sampling method. Even if minor, this process could lead to certain biases and imbalances. 
Similarly we observe that alternating between image-image and caption-caption similarity can improve \textit{CLIP-NESt}; presumably by generating more diverse image-caption pairs.
Furthermore, our results show that feature-based negative sampling, including both \textit{CSt} and \textit{NC}, can surpass random negative sampling, due to their ability to generate hard negative OOC pairs, which more accurately reflect real-world misinformation.
Finally, all multimodal OOC-based Synthetic Misinformers outperform their unimodal counterparts (image-only and text-only) and therefore do not suffer from a unimodal bias. 

Shifting to NEI Misinformers, we see that both \textit{R-NESt} and \textit{CLIP-NESt} surpass \textit{MEIR} \cite{sabir2018deep}. 
The methodology could be a contributing factor but we should also consider the difference in data scale. \textit{MEIR} consists of only 82,156 truthful and 57,940 falsified instances compared to the approximately 1,2M data falsified points produced by other Misinformers, which may be inadequate to correctly train the Detector. 
We also observe that the proposed \textit{CLIP-NESt-alt} surpasses all other multimodal NEI and OOC methods; achieving 56.9\% accuracy.
Furthermore, combining R-NESt with CSt-alt achieved the same score but using NC/I-T in conjunction with R-NESt did not perform as well since NC/I-T consists of 226,564 samples and required heavily down-sampling truthful and R-NESt samples (1M each), resulting in a notably smaller dataset.
Finally we observe that our hybrid \textit{CLIP-NESt-alt + CSt-alt} method proved capable of achieving the highest multimodal accuracy (58.1\%). 

Nevertheless, we also recognize certain problems and limitations.
First, even the best performing methods have trouble accurately identifying the falsified pairs, scoring lower than 50\% in terms of \texttt{Falsified} Hit Rate (Sensitivity) while having high \texttt{Truthful} Hit Rate scores (Specificity). 
This indicates that the task of multimodal misinformation detection is significantly more challenging than previous studies suggested, e.g. showcasing scores higher than 88\% on the COSMOS dataset \cite{aneja2021cosmos, akgul2021cosmos, alkaddour2022sentiment} while using a problematic evaluation protocol 
(discussed in Section \ref{sec:eval_protocol}). 
Finally, we found that unimodal text-only methods, such as \textit{CLIP-NESt-C} and \textit{R-NESt}, \textit{R-NESt + CSt-alt}, can outperform their multimodal counterparts - with the latter scoring 59.1\% - on a supposedly multimodal task. This suggests the existence of a unimodal bias in the dataset, which needs to be addressed in future studies.

\begin{table*}
\centering
  \caption{Comparative study between numerous Synthetic Misinformer methods evaluated on the COSMOS benchmark. We use the DT-Transformer and the CLIP ViT L/14 encoder. We report the Accuracy of unimodal (image-only, text-only) and multimodal Detectors as well as the multimodal Hit Rate per class (Truthful and Falsified pairs) or Specificity and Sensitivity respectively. \textbf{Bold} denotes the overall highest accuracy while \underline{underline} denotes the highest multimodal accuracy. 
  }
  
  \label{tab:comparative}
  \begin{tabular}{llc|c|ccc}
    \toprule

    & & \textbf{Image-only} & \textbf{Text-only} & & \textbf{Multimodal} & \\
    
     \textbf{Type} &
     \textbf{Synthetic Misinformer} & 
     Accuracy &      
     Accuracy & 
     Accuracy & 
     Truthful &
     Falsified 
     \\
     \midrule

    \multirow{10}{*}{OOC} & RS-C & 50.0 & 50.0 & 52.2 & 92.4 & 12.0 \\
    & RSt-C & 50.0 & 50.0 & 53.8 & 93.5 & 14.0 \\
    & RSt-I & 50.2 & 50.0 & 52.3 & 96.6 & 8.0 \\
    & RSt-alt & 50.0 & 50.0 & 53.9 & 92.0 & 15.9 \\

   &  CSt-C & 50.0 & 48.5 & 53.1 & 70.7 & 35.5 \\
    & CSt-I & 49.5 & 50.0 & 54.7 & 83.4 & 25.9 \\
    & CSt-alt & 52.3 & 51.5 & 55.0 & 76.6 & 33.4 \\

    & NC/Bal \cite{luo2021newsclippings} & 50.0 & 50.0 & 54.1 & 52.8 & 55.3 \\
    & NC/T-T \cite{luo2021newsclippings} & 48.8 & 50.0 & 53.7 & 66.1 & 41.3 \\
    & NC/I-T \cite{luo2021newsclippings} & 49.2 & 50.0 & 55.4 & 84.6 & 26.2 \\

    \midrule

    \multirow{5}{*}{NEI} & MEIR \cite{sabir2018deep} & 49.4 & 50.5 & 53.9 & 90.5 & 17.3 \\
    & R-NESt & 50.1 & 58.5 & 55.7 & 90.5 & 20.8 \\
    & CLIP-NESt-C & 50.0 & 58.1 & 55.3 & 75.1	& 35.5 \\
    & CLIP-NESt-I & 50.0 & 54.1 & 55.2 & 81.2 & 29.3 \\
    & CLIP-NESt-alt & 50.0 & 56.7 & 56.9 & 81.7 & 32.2 \\

    \midrule
    \multirow{3 }{*}{Hybrid} & R-NESt + CSt-alt & 51.8 & \textbf{59.1} & 56.9 & 89.3 & 24.5 \\
    & R-NESt + NC/I-T & 51.5 & 57.2 & 54.4 & 90.5 & 18.1 \\
    & CLIP-NESt-alt + CSt-alt & 51.3 & 52.8 & \underline{58.1} & 74.4 & 41.9 \\

    \bottomrule
  \end{tabular}
\end{table*}

\section{Conclusions}

In this study we address the task of multimodal misinformation detection (MMD). More specifically, we examine and compare multiple methods that generate training data (Synthetic Misinformers) for MMD, either out-of-context image-text pairs (OOC) or cross-modal named entity inconsistencies (NEI).
We perform a comparative study and evaluate all Synthetic Misinformers on the COSMOS benchmark; consisting of real-world multimodal misinformation. 
The comparative study illustrated that NEI methods tend to - on average - outperform OOC methods on the COSMOS benchmark. Moreover, our proposed \textit{CLIP-NESt-alt} method reached the highest multimodal accuracy (56.9\%) among NEI and OOC methods; having a 2.15\% and 2.7\% advantage over the next best performing method from each respectively. 
Furthermore, we hypothesized that real-world misinformation is not solely captured by OOC or NEI instances separately but instead necessitate both.
This is validated by the proposed hybrid approach (\textit{CLIP-NESt-alt + CSt-alt}) achieving the highest multimodal accuracy (58.1\%); showing a 2.47\% improvement over \textit{CLIP-NESt-alt}.

Nevertheless, low Sensitivity scores (Table \ref{tab:comparative}) indicate that - under the corrected evaluation protocol - MMD is a significantly more challenging task than previous works suggested \cite{aneja2021cosmos} and extensive further research is essential.
Future studies could consider the integration of external evidence \cite{abdelnabi2022open, glockner2022missing} or knowledge graphs \cite{liu2022comparative} 
not only to improve detection accuracy but also to develop new Synthetic Misinformers that generate more realistic synthetic training data, and as a result produce better Detectors.
Moreover, experimentation with different modality fusion techniques can further improve performance \cite{kumari2021amfb, yu2022bcmf}.
Furthermore, our empirical results showed that NEI Misinformers tend to introduce a unimodal bias, leading to unimodal Detectors competing or even outperforming multimodal ones.
Named entity manipulations could create certain linguistic patterns, biases or shortcuts that render the visual information less important.
Future studies could explore developing methods for generating de-biased NEI or learning strategies for reducing unimodal bias \cite{cadene2019rubi}.
Moreover, task-specific modality fusion methods could potentially help mitigate this challenge \cite{li2021entity}. 
Finally, the unimodal bias may not lie with the training process but with the evaluation dataset. The COSMOS benchmark was not collected with certain criteria in place to explicitly make it difficult for unimodal architectures.
Future studies could explore and define relevant rules and criteria for collecting a more robust real-world MMD benchmark.

\section{Acknowledgments}

This work is partially funded by the project ``vera.ai: VERification Assisted by Artificial Intelligence'' under grant agreement no. 101070093.

\bibliographystyle{unsrt}  
\bibliography{references}

\begin{thebibliography}{10}

\bibitem{mridha2021comprehensive}
Muhammad~F Mridha, Ashfia~Jannat Keya, Md~Abdul Hamid, Muhammad~Mostafa
  Monowar, and Md~Saifur Rahman.
\newblock A comprehensive review on fake news detection with deep learning.
\newblock {\em IEEE Access}, 9:156151--156170, 2021.

\bibitem{rana2022deepfake}
Md~Shohel Rana, Mohammad~Nur Nobi, Beddhu Murali, and Andrew~H Sung.
\newblock Deepfake detection: A systematic literature review.
\newblock {\em IEEE Access}, 2022.

\bibitem{li2020picture}
Yiyi Li and Ying Xie.
\newblock Is a picture worth a thousand words? an empirical study of image
  content and social media engagement.
\newblock {\em Journal of Marketing Research}, 57(1):1--19, 2020.

\bibitem{newman2012nonprobative}
Eryn~J Newman, Maryanne Garry, Daniel~M Bernstein, Justin Kantner, and
  D~Stephen Lindsay.
\newblock Nonprobative photographs (or words) inflate truthiness.
\newblock {\em Psychonomic Bulletin \& Review}, 19:969--974, 2012.

\bibitem{alam2021survey}
Firoj Alam, Stefano Cresci, Tanmoy Chakraborty, Fabrizio Silvestri, Dimiter
  Dimitrov, Giovanni Da~San Martino, Shaden Shaar, Hamed Firooz, and Preslav
  Nakov.
\newblock A survey on multimodal disinformation detection.
\newblock {\em arXiv preprint arXiv:2103.12541}, 2021.

\bibitem{aneja2021cosmos}
Shivangi Aneja, Chris Bregler, and Matthias Nie{\ss}ner.
\newblock Cosmos: Catching out-of-context misinformation with self-supervised
  learning.
\newblock {\em arXiv preprint arXiv:2101.06278}, 2021.

\bibitem{muller2020multimodal}
Eric M{\"u}ller-Budack, Jonas Theiner, Sebastian Diering, Maximilian Idahl, and
  Ralph Ewerth.
\newblock Multimodal analytics for real-world news using measures of
  cross-modal entity consistency.
\newblock In {\em Proceedings of the 2020 International Conference on
  Multimedia Retrieval}, pages 16--25, 2020.

\bibitem{jaiswal2017multimedia}
Ayush Jaiswal, Ekraam Sabir, Wael AbdAlmageed, and Premkumar Natarajan.
\newblock Multimedia semantic integrity assessment using joint embedding of
  images and text.
\newblock In {\em Proceedings of the 25th ACM international conference on
  Multimedia}, pages 1465--1471, 2017.

\bibitem{luo2021newsclippings}
Grace Luo, Trevor Darrell, and Anna Rohrbach.
\newblock Newsclippings: Automatic generation of out-of-context multimodal
  media.
\newblock {\em arXiv preprint arXiv:2104.05893}, 2021.

\bibitem{biamby2021twitter}
Giscard Biamby, Grace Luo, Trevor Darrell, and Anna Rohrbach.
\newblock Twitter-comms: Detecting climate, covid, and military multimodal
  misinformation.
\newblock {\em arXiv preprint arXiv:2112.08594}, 2021.

\bibitem{sabir2018deep}
Ekraam Sabir, Wael AbdAlmageed, Yue Wu, and Prem Natarajan.
\newblock Deep multimodal image-repurposing detection.
\newblock In {\em Proceedings of the 26th ACM international conference on
  Multimedia}, pages 1337--1345, 2018.

\bibitem{liu2020visual}
Fuxiao Liu, Yinghan Wang, Tianlu Wang, and Vicente Ordonez.
\newblock Visual news: Benchmark and challenges in news image captioning.
\newblock {\em arXiv preprint arXiv:2010.03743}, 2020.

\bibitem{radford2021learning}
Alec Radford, Jong~Wook Kim, Chris Hallacy, Aditya Ramesh, Gabriel Goh,
  Sandhini Agarwal, Girish Sastry, Amanda Askell, Pamela Mishkin, Jack Clark,
  et~al.
\newblock Learning transferable visual models from natural language
  supervision.
\newblock In {\em International conference on machine learning}, pages
  8748--8763. PMLR, 2021.

\bibitem{akgul2021cosmos}
Tankut Akgul, Tugce~Erkilic Civelek, Deniz Ugur, and Ali~C Begen.
\newblock Cosmos on steroids: a cheap detector for cheapfakes.
\newblock In {\em Proceedings of the 12th ACM Multimedia Systems Conference},
  pages 327--331, 2021.

\bibitem{alkaddour2022sentiment}
Muhannad Alkaddour, Abhinav Dhall, Usman Tariq, Hasan Al~Nashash, and Fares
  Al-Shargie.
\newblock Sentiment-aware classifier for out-of-context caption detection.
\newblock In {\em Proceedings of the 30th ACM International Conference on
  Multimedia}, pages 7180--7184, 2022.

\bibitem{abdelnabi2022open}
Sahar Abdelnabi, Rakibul Hasan, and Mario Fritz.
\newblock Open-domain, content-based, multi-modal fact-checking of
  out-of-context images via online resources.
\newblock In {\em Proceedings of the IEEE/CVF Conference on Computer Vision and
  Pattern Recognition}, pages 14940--14949, 2022.

\bibitem{devlin2018bert}
Jacob Devlin, Ming-Wei Chang, Kenton Lee, and Kristina Toutanova.
\newblock Bert: Pre-training of deep bidirectional transformers for language
  understanding.
\newblock {\em arXiv preprint arXiv:1810.04805}, 2018.

\bibitem{bojanowski2017enriching}
Piotr Bojanowski, Edouard Grave, Armand Joulin, and Tomas Mikolov.
\newblock Enriching word vectors with subword information.
\newblock {\em Transactions of the association for computational linguistics},
  5:135--146, 2017.

\bibitem{huang2022text}
Mingzhen Huang, Shan Jia, Ming-Ching Chang, and Siwei Lyu.
\newblock Text-image de-contextualization detection using vision-language
  models.
\newblock In {\em ICASSP 2022-2022 IEEE International Conference on Acoustics,
  Speech and Signal Processing (ICASSP)}, pages 8967--8971. IEEE, 2022.

\bibitem{wortsman2022robust}
Mitchell Wortsman, Gabriel Ilharco, Jong~Wook Kim, Mike Li, Simon Kornblith,
  Rebecca Roelofs, Raphael~Gontijo Lopes, Hannaneh Hajishirzi, Ali Farhadi,
  Hongseok Namkoong, et~al.
\newblock Robust fine-tuning of zero-shot models.
\newblock In {\em Proceedings of the IEEE/CVF Conference on Computer Vision and
  Pattern Recognition}, pages 7959--7971, 2022.

\bibitem{vaswani2017attention}
Ashish Vaswani, Noam Shazeer, Niki Parmar, Jakob Uszkoreit, Llion Jones,
  Aidan~N Gomez, {\L}ukasz Kaiser, and Illia Polosukhin.
\newblock Attention is all you need.
\newblock {\em Advances in neural information processing systems}, 30, 2017.

\bibitem{aneja2021mmsys}
Shivangi Aneja, Cise Midoglu, Duc-Tien Dang-Nguyen, Michael~Alexander Riegler,
  Paal Halvorsen, Matthias Nie{\ss}ner, Balu Adsumilli, and Chris Bregler.
\newblock Mmsys' 21 grand challenge on detecting cheapfakes.
\newblock {\em arXiv preprint arXiv:2107.05297}, 2021.

\bibitem{glockner2022missing}
Max Glockner, Yufang Hou, and Iryna Gurevych.
\newblock Missing counter-evidence renders nlp fact-checking unrealistic for
  misinformation.
\newblock {\em arXiv preprint arXiv:2210.13865}, 2022.

\bibitem{liu2022comparative}
Lihui Liu, Houxiang Ji, Jiejun Xu, and Hanghang Tong.
\newblock Comparative reasoning for knowledge graph fact checking.
\newblock In {\em 2022 IEEE International Conference on Big Data (Big Data)},
  pages 2309--2312. IEEE, 2022.

\bibitem{kumari2021amfb}
Rina Kumari and Asif Ekbal.
\newblock Amfb: Attention based multimodal factorized bilinear pooling for
  multimodal fake news detection.
\newblock {\em Expert Systems with Applications}, 184:115412, 2021.

\bibitem{yu2022bcmf}
Chuanming Yu, Yinxue Ma, Lu~An, and Gang Li.
\newblock Bcmf: A bidirectional cross-modal fusion model for fake news
  detection.
\newblock {\em Information Processing \& Management}, 59(5):103063, 2022.

\bibitem{cadene2019rubi}
Remi Cadene, Corentin Dancette, Matthieu Cord, Devi Parikh, et~al.
\newblock Rubi: Reducing unimodal biases for visual question answering.
\newblock {\em Advances in neural information processing systems}, 32, 2019.

\bibitem{li2021entity}
Peiguang Li, Xian Sun, Hongfeng Yu, Yu~Tian, Fanglong Yao, and Guangluan Xu.
\newblock Entity-oriented multi-modal alignment and fusion network for fake
  news detection.
\newblock {\em IEEE Transactions on Multimedia}, 24:3455--3468, 2021.

\end{thebibliography}

\end{document}